\documentstyle[12pt,aasms4]{article}
\catcode`\@=11
\def\bibstyle@aa{\bibpunct{(}{)}{;}{a}{}{,}}
\def\bibstyle@pass{\bibpunct{(}{)}{;}{a}{,}{,}}
\def\bibstyle@anngeo{\bibpunct{(}{)}{;}{a}{,}{,}}
\def\bibstyle@agsm{\bibpunct{(}{)}{,}{a}{}{,}\gdef\harvardand{\&}}
\def\bibstyle@kluwer{\bibpunct{(}{)}{,}{a}{}{,}\gdef\harvardand{\&}}
\def\bibstyle@dcu{\bibpunct{(}{)}{;}{a}{;}{,}\gdef\harvardand{and}}
\def\bibstyle@agu{\bibpunct{[}{]}{;}{a}{,}{,}}
\def\bibstyle@nlinproc{\bibpunct{(}{)}{;}{a}{,}{,}}
\def\bibstyle@cospar{\bibpunct{/}{/}{,}{n}{}{}%
     \gdef\@biblabel##1{##1.}}
\def\bibstyle@esa{\bibpunct{(}{)}{,}{n}{}{}%
     \gdef\@biblabel##1{##1.\hspace{1em}}%
     \gdef\@cite##1##2##3{\@citebegin Ref.~##1\if@tempswa,
          ##3\fi\@citeend}}
\def\bibstyle@nature{\bibpunct{}{}{,}{n}{}{}%
     \gdef\@biblabel##1{##1.}%
     \gdef\@cite##1##2##3{\unskip\mbox{$^{##1}$}}}
\def\bibstyle@plain{\bibpunct{[}{]}{,}{n}{}{}}
\let\bibstyle@alpha=\bibstyle@plain
\let\bibstyle@abbrv=\bibstyle@plain
\let\bibstyle@unsrt=\bibstyle@plain
\def\@cite#1#2#3{\if@tempswa\@citebegin\if#2\@empty\else#2 \fi
        #1\if#3\@empty\else, #3\fi\@citeend\else#1\fi}
\def\@citenum#1#2#3{\@citebegin\if@tempswa\if#2\@empty\else#2 \fi\fi
   #1\if#3\@empty\else, #3\fi\@citeend}
\def\@citexnum[#1][#2]#3{\if@filesw\immediate\write
      \@auxout{\string\citation{#3}}\fi
  \let\@citea\@empty
  \@cite{\@for\@citeb:=#3\do
    {\@citea\def\@citea{\@citesep\penalty\@m\ }%
     \def\@tempa##1##2\@nil{\edef\@citeb{\if##1\space##2\else##1##2\fi}}%
     \expandafter\@tempa\@citeb\@nil
     \@ifundefined{b@\@citeb}{%
       {\reset@font\bf ?}\@warning
       {Citation `\@citeb' on page \thepage \space undefined}}%
     \hbox{\csname b@\@citeb\endcsname}}}{#1}{#2}}
\def\@citex[#1][#2]#3{\let\@citea\@empty
  \@cite{\let\@citenm\@empty
    \@for\@citeb:=#3\do
    {\edef\@citeb{\expandafter\@iden\@citeb}%
     \if@filesw\immediate\write\@auxout{\string\citation{\@citeb}}\fi
     \@ifundefined{b@\@citeb}{\@citea%
       {\reset@font\bf ?}\@warning
       {Citation `\@citeb' on page \thepage \space undefined}}%
     {\let\@citemm=\@citenm
     \@cite@parse{\@citeb}%
     \if@tempswa
       \ifx\@citemm\@citenm\@yrsep\else\@citea{\@citenm}\@auyrsep\fi
       \ \@citedt \def\@citea{\@citesep\ }%
     \else
       \ifx\@citemm\@citenm, \@citedt\else\@citea{\@citenm}
           \@citebegin\@citedt\fi
       \def\@citea{\@citeend\@citesep\ }%
     \fi}}\if@tempswa\else\@citeend\fi}{#1}{#2}}
\def\@biblabel#1{\hfill}
\def\@biblabelnum#1{[#1]}
\def\@bibsetnum#1{\settowidth\labelwidth{\@biblabel{#1}}%
   \leftmargin\labelwidth \advance\leftmargin\labelsep
}
\def\@bibsetup#1{\leftmargin=1em\itemindent=-\leftmargin}
\def\@citebegin{(} \def\@citeend{)} \def\@citesep{;}
\def\@auyrsep{} \def\@yrsep{}
\def\bibstyle#1{\@ifundefined{bibstyle@#1}{\relax}
     {\csname bibstyle@#1\endcsname}}
\let\ori@document=\document
\def\document{\ori@document\global\let\bibstyle=\@gobble}
\def\bibpunct#1#2#3#4#5#6{\gdef\@citebegin{#1}\gdef\@citeend{#2}\gdef
   \@citesep{#3}\ifx #4n\global\let\@bibsetup=\@bibsetnum
   \global\let\@citex=\@citexnum
   \global\let\@biblabel=\@biblabelnum
   \global\let\@cite=\@citenum\fi
   \gdef\@auyrsep{#5}\gdef\@yrsep{#6}}
\newif\ifNAT@full\NAT@fullfalse
\def\cite{\@ifstar{\NAT@fulltrue\@citee}{\NAT@fullfalse\@citee}}
\def\@citee{\@ifnextchar [{\@tempswatrue\@citex@}{\@tempswafalse
    \@citex@[]}}
\def\@citex@[#1]{\@ifnextchar [{\@citex[#1]}{\@citex[][#1]}}
\newcommand{\citeauthor}[1]{\if@filesw\immediate\write
     \@auxout{\string\citation{#1}}\fi
\ifx\@citex\@citexnum
       {\reset@font\bf(author?)}\@warning
       {Cannot use \protect\citeauthor
         ^^J
        with numerical citations}\else
     \@ifundefined{b@#1}{%
       {\reset@font\bf ?}\@warning
       {Citation `#1' on page \thepage \space undefined}}%
       {\@cite@parse{#1}{\@citenm}}\fi}
\newcommand{\citeyear}[1]{\if@filesw\immediate\write
   \@auxout{\string\citation{#1}}\fi
\ifx\@citex\@citexnum
       {\reset@font\bf(year?)}\@warning
       {Cannot use \protect\citeyear
         ^^J
        with numerical citations}\else
     \@ifundefined{b@#1}{%
       {\reset@font\bf ?}\@warning
       {Citation `#1' on page \thepage \space undefined}}%
       {\@cite@parse{#1}\@citedt}\fi}
\newcommand{\citefullauthor}[1]{\if@filesw\immediate\write
     \@auxout{\string\citation{#1}}\fi
\ifx\@citex\@citexnum
       {\reset@font\bf(author?)}\@warning
       {Cannot use \protect\citeauthor
         ^^J
        with numerical citations}\else
     \@ifundefined{b@#1}{%
       {\reset@font\bf ?}\@warning
       {Citation `#1' on page \thepage \space undefined}}%
       {\@cite@parse{#1}{\@citefull}}\fi}
\def\@cite@parse#1{{%
      \@ifundefined{documentclass}
       {\let\prm=\relax\let\psf=\relax\let\ptt=\relax\let\pbf=\relax
        \let\psl=\relax\let\psc=\relax\let\pit=\relax\let\pem=\relax
        \let\prmfamily=\relax\let\psffamily=\relax\let\pttfamily=\relax
        \let\pbfseries=\relax\let\pslshape=\relax\let\pscshape=\relax
        \let\pitshape=\relax\let\pmdseries=\relax\let\pupshape=\relax
        \let\pc=\relax \let\pd=\relax \let\pb=\relax}
       {\let\protect=\@unexpandable@protect}%
     \xdef\@tempa{\csname b@#1\endcsname\relax}}%
     \expandafter\@citez\@tempa()()\@nil
     \ifNAT@full\let\@citenm\@citefull\fi}
\def\@citez#1(#2)#3()#4\@nil{\gdef\@citenm{#1}\gdef\@citedt{#2}%
  \ifx#3\relax\gdef\@citefull{#1}\else\gdef\@citefull{#3}\fi
  \if!#2!\expandafter\@citeapalk#1\@nil\fi}
\def\@citeapalk#1, #2\@nil{\gdef\@citenm{#1}\gdef\@citedt{#2}%
   \gdef\@citefull{#1}}
\newcommand{\citeauthoryear}[3]{\ifx#3\relax #1(#2)#1\else #2(#3)#1\fi}
\newcommand{\citestarts}{\@citebegin}
\newcommand{\citeends}{\@citeend}

\def\harvarditem{\@ifnextchar[{\@harvarditem}{\@harvarditem[\@empty]}}
\def\@harvarditem[#1]#2#3#4{\if!#1!\bibitem[#2(#3)]{#4}\else
  \bibitem[#1(#3)#2]{#4}\fi }
\newcommand{\harvardleft}{\@citebegin}
\newcommand{\harvardright}{\@citeend}
\newcommand{\harvardyearleft}{\@citebegin}
\newcommand{\harvardyearright}{\@citeend}
\def\harvardand{and}

\@ifundefined{chapter}{\def\bibsection{\section*{\refname
   \@mkboth{\uppercase{\refname}}{\uppercase{\refname}}}}}{\def
   \bibsection{\chapter*{\bibname
   \@mkboth{\uppercase{\bibname}}{\uppercase{\bibname}}}}}

\@ifundefined{reset@font}{\let\reset@font=\relax}{}

\catcode`\@=12

\newcommand\rd{{\rm d}}
\newcommand\eref[1]{(\ref{#1})}

\newcommand\rme {{\rm e }}
\newcommand\rmi {{\rm i }}
\begin{document}
\slugcomment{Accepted for publication in {\em The Astrophysical Journal}}
\title{Frequency Analysis of Reflex Velocities of Stars with Planets}

\author{Maciej Konacki and Andrzej J. Maciejewski
\affil{Toru\'n Centre for Astronomy, 
Nicolaus Copernicus University,\\
87-100 Toru\'n, Gagarina 11, Poland;
kmc@astri.uni.torun.pl
}
}

\begin{abstract}
Since it has become possible to discovery planets orbiting nearby solar-type 
stars through very precise Doppler-shift measurements, the role of 
methods used to analyze such observations has grown significantly. 
The widely employed model-dependent approach based on the least-squares 
fit of the Keplerian motion to the radial-velocity variations can be, as we
show, unsatisfactory. Thus, in this paper, we propose a new method that may be
easily and successfully applied to the Doppler-shift measurements. 
This method allows us to analyze the data without assuming any specific model and
yet to extract all significant features of the observations. This very
simple idea, based on the subsequent subtraction of all harmonic components
from the data, can be easily  implemented. We show that our method
can be used to analyze real 16 Cygni B Doppler-shift observations with a 
surprising but correct result which is substantially different from that based 
on the least-squares fit of a Keplerian orbit. Namely, using  frequency 
analysis we show that with the current accuracy of this star's observations 
it is not possible to determine the value of the orbital eccentricity which
is claimed to be as high as 0.6.
\end{abstract}
\section{Introduction}

Recent improvements in the long-term precision of Doppler-shift measurements
\cite[]{Marcy:96c::} resulted in several spectacular detections of 
planetary companions to solar-type stars \cite[for review see the paper of][]
{Marcy:97::}. As such discoveries supply indirect evidence of 
the existence of extra-solar planets,  other explanations 
of observed 
radial-velocity variations appeared, e.g., stellar pulsations \cite[]{Gray:97::}. 
The most recent results, however, show that only a planetary hypothesis is  
acceptable \cite[]{Marcy:98::,Gray:98::}.  

The usual procedure showing that there exists a planet around s star consits of  
a direct least-square fit of Keplerian model to  the observations. 
It always gives   certain values of orbital parameters and their formal errors.  
In the case of  `good' data this is the best and the quickest way to obtain relable  
results. However, in the case of spare data with big   
errors one has to prove  that  the least-squares method can be used  and that  
the obtained  parameter values   and their errors are good estimates of the real 
values. This is a difficult  and time consuming task. Without doubt, we have this 
situation  with the  Doppler observation of extra-solar planets.  We present an 
analysis of this problem and  we show that the eccentricity of the fitted orbit is  
a very sensitive parameter and, in some cases, its value   and   error  given by
the least-squares method are not correct. 

The aim of this paper is to show how the mentioned problem can be solved in practice.
Namely, we propose a method  that can be very useful for analyzing radial-velocity 
variations.  It is based on a simple idea involving the subsequent subtraction of 
periodic components from the data. This approach allows us to analyze the observations without assuming any specific
model describing the system  behavior (like Keplerian motion or stellar
pulsations). After the determination of all significant components of the
data, it remains to be decided which process is responsible for what we observe
and whether it is possible to choose only one.

The plan of this paper is  as follows. In Section 2 we  analyse observations  of
16 Cygni B and we explain why  the standard least-squere fit does not give 
reliable estimates of parameters and their errors. In Section 3 we analyze
analytically the  Keplerian motion of the system `a star with one planet' in
order to learn how its motion modulates the observed star radial-velocities. We
investigate  mainly the spectral properties of the motion which are essential
for our method. In Section 4 we develop a simple numerical technique which can
be used to extract all the information we need to compare with the results from
Section 3. In Section 5, we perform a numerical
test of the method using  simulated radial-velocity variations with the orbital
parameters of 70 Vir  \cite[]{Marcy:96b::}. In section 6, we discuss the
application of the method to finding the eccentricity of 16 Cygni B
\cite[]{Cochran:97::}. 

\section{Least-squares analysis of radial velocities of 16 Cygni B}

In order to determine the quality and reliability of the least-squares
fit of a Keplerian orbit to the radial velocity measurements of 16 Cygni B
performed by \cite{Cochran:97::} we analyzed the topology of the $\chi^2$ minimum
on the $(a,e)$ plane, where $a$ is the semi-major axis and $e$ is the
eccentricity. To this end, we used the real observations published in
\cite{Cochran:97::} and the Levenberg-Marquard method to 
solve the nonlinear least-squares  problem (in our case a fit of a Keplerian
orbit) which cannot be linearized (as this is the case for 16 Cygni B
observations, as we show below). 
We took the necessary FORTRAN code from MINPACK 
library \cite[]{minpack:::}. As  the semi-major axis and the eccentricity  
are the most crucial parameters of the model, we compute the behavior of 
\[
\chi^2 = \sum_i\frac{(v^{i}_{modeled} - v^{i}_{observed} )^2}{\sigma_i^2} 
\]
on the $(e,a)$ plane. For a given point $(a,e)$ other parameters $(P,T_p,\omega)$ are 
always chosen to correspond to the global minimum of 
$\chi^2$, i.e., for  fixed values of parameters $(a,e)$ we make a series of fits with 
initial values 
of the elements $(P,T_p,\omega)$ covering their whole range and we take the 
smallest value of $\chi^2$. In this way
we get the behavior of $\chi^2$ on the $(e,a)$ plane, see Figure \ref{16cygbplan}.
As one can see the confidence levels $1\sigma,2\sigma,3\sigma$ of the 
parameters $(e,a)$ bound a large region of the parameters' plane. This shows
that $e$ is very weakly determined with the data available for 16 Cygni B.
At this point we make a well known but important remark. Namely, as one can see 
in Figure \ref{16cygbplan}, the confidence lines
do not determine ellipses as it should be if we could use
$1\sigma,2\sigma,3\sigma$ levels to determine the errors of $a$ and $e$ 
\cite[]{Press:92::}.
It also means that the errors of the parameters obtained from the 
linearized (or not) least-squares fit cannot be treated as the correct
estimates of the parameters' accuracy. Thus, for example, the value
0.082 from the paper of \cite{Cochran:97::} has little to do with
the accuracy of the determined eccentricity. In fact, using nonlinear
Levenberg-Marquard least-squares method we find a better
solution for the same data of 16 Cygni B (this solution is marked with
a filled square in Figure \ref{16cygbplan}; the filled star indicates the 
orbital solution 
found by \cite{Cochran:97::}). One can compare the $\chi^2$ behavior
with a very similar picture obtained by means of the bootstrap method 
\cite[]{Press:92::}, see Figure \ref{bootstrap}. Clearly the distribution 
of $a$ and $e$ is not normal. Moreover, from this distribution we learn that
any value of the eccentricity within the interval approximately $(0.25,1.0)$
is probable at the $3\sigma$ level. 

To sum up, one can always use the least-squares method 
to obtain the orbital parameters. This method always gives a solution
but sometimes the obtained parameters together with its errors 
have little to do with the real ones, whatever they are.  
However, the process of checking if the obtained result is correct is time 
consuming, and thus it seems resonable to find a simple method which can 
give the correct answer quickly. We address this problem in Sections 3 and 4.

\section{Theoretical Background of Frequency Analysis}

Let us assume that we have a planetary system consisting of a star and one 
planet.
Choosing the reference frame placed in the center of mass of this system
(barycentric system) with the $Z$-th axis directed from the observer, we
can calculate the $Z$-th coordinate of the star from the following equation
\begin{equation}
\label{t:1::}
 Z_{\star}(t) = - \frac{m}{m_{\star}}Z(t),
\end{equation}
where $m_{\star}$ is the mass of the star, $m$ is the mass
of the planet and $Z$ is its $Z$-th coordinate. Next, because of
the Keplerian motion of the system, we have
\begin{equation}
\label{t:2::}
 Z(t) = a \sin {i}
 \left( \sin\omega\left(\cos E - e\right)
 + \cos\omega\sqrt{1 - e^2}\sin E \right),
\end{equation}
where $a, i, \omega, e$ are Keplerian elements of the planet (semi-major axis, 
inclination, longitude of periastron and eccentricity) and $E$ is its eccentric 
anomaly that can be calculated from the Kepler equation
\begin{equation}
\label{t:3::}
  E - e\sin E = M, \qquad M = n(t - T_{p}), \qquad n = \frac{2\pi}{P},
\end{equation}
where $M$ is the mean anomaly, $P$ is the orbital period of the planet 
and $T_{p}$ is the time of periastron. It was shown by \cite{Konacki:96::} 
that  function \eref{t:1::} can be expanded in the following series
\begin{equation}
\label{t:4::}
Z_{\star}(t) = -\frac{m}{m_{\star}}\left(Z_{0}  +
                \sum_{k=-\infty}^{k=+\infty}\!\!\!\!{}^{'}\,
                Z_{k}\rme^{\rmi kn(t-t_0)}\right),
\end{equation}
where
\[
 Z_{k} = \frac{1}{2k}\left\{
        C[ J_{k - 1}(ke) - J_{k + 1}(ke) ] +
 \rmi S[ J_{k - 1}(ke) + J_{k + 1}(ke) ] \right\}
 \rme^{-\rmi knT_{p}},
\]
and
\[
Z_{0} = -\frac{3}{2}ea\sin i\sin\omega, \quad
C = a\sin i\sin\omega, \quad
S = -a\sqrt{1 - {e}^2}\sin i\cos\omega.
\]
In the above $J_n(z)$ is a Bessel function of argument $z$.
 
This expansion has a very important property---terms corresponding
to successive harmonics have decreasing amplitudes. In this way, the term with
the  frequency $n$ has a larger amplitude than that with frequency $2n$, which
has larger amplitude than that with frequency $3n$, etc. Generally, it is
possible to prove the following inequality \cite[]{Konacki:96::} 
\begin{equation}
\label{t:5::}
  {\cal A}_k(Z) = \frac{\left|Z_{k+1}\right|}
                {\left|Z_{k}\right|} =
         \frac{k}{k+1}\Psi_k(e,\omega) < 1, 
\end{equation}
for each $e \in (0,1)$ and $\omega \in [0,2\pi)$; in \eref{t:5::} we denote
\begin{equation}
\label{t:5::1}
\Psi_k(e,\omega)  =  \sqrt{\frac{ \beta^2\tan^2\omega [J_{k+1}'((k+1)e)]^2 
                                        + [J_{k+1}((k+1)e)]^2 }
                { \beta^2\tan^2\omega[J_{k}'(ke)]^2 
                                       + [J_{k}(ke)]^2} },
\end{equation}
and $\beta^{2} = e^2/(1-e^2)$. 
When $e = 0$ only the first term with  frequency $n$ has a non-zero 
amplitude.
 
These considerations allow us to state the following: Keplerian motion
of the planet orbiting a star leads to specific changes in the $Z$-th
coordinate of a star. These changes have a very characteristic 
spectrum in which the term with frequency equal to the mean motion of the planet 
$n$ is dominant and amplitudes of subsequent harmonics (with frequencies 
$2n, 3n, \dots$) decrease strictly monotonically.

The expansion of the Keplerian motion we show above can be easily applied to 
calculate the  radial velocity of the star  resulting from its motion in the 
system. Having the expansion of the $Z$-th coordinate of the star, we derive 
its radial-velocity by differentiating \eref{t:4::} with respect to  time 
\begin{equation}
\label{t:6::}
 v_r(t) = -\frac{\rd Z_{\star}(t)}{\rd t} = n {\frac{m}{m_{\star}}}
 \sum_{k=-\infty}^{k=+\infty}\!\!\!\!{}^{'}\,
         \rmi k Z_{k}{\rme^{\rmi kn(t-t_0)}}.
\end{equation}
As it can be easily noticed, the above expansion has the same feature as the 
expansion of the $Z$-th coordinate. In fact, the ratio  ${{\cal A}_{k}(v_r)}$
of two successive harmonics is always less than 1, since we have 
\cite[]{Konacki:96::}
\begin{equation}
\label{t:8::}
  {{\cal A}_k (v_r)} = \frac{\left|Z_{k+1}(k+1)n\right|}
                {\left|Z_{k}kn\right|} =\frac{k+1}{k}{\cal A}_k(Z) = 
\Psi_k (e,\omega) <1.
\end{equation}
Thus, all useful properties of the radial velocity expansion can be derived 
from that of the $Z$-th coordinate. 

In summary, if we observe radial-velocity variations that are of planetary
origin then in their spectra we will detect the basic frequency corresponding
to the planet orbital period and its harmonics with 
monotonically decreasing amplitudes.
The ratios of successive harmonics depend on the value of the eccentricity and the 
longitude of periastron. It means that in the case of
a circular orbit we will be able to notice only one periodic term (which
is obvious) and with increasing eccentricity the number of detectable
harmonics will increase. Finally, we should mention that, because of the
finite accuracy of our observations, we can only detect a few  higher
harmonics. Thus, from the observational point of view,  expansion \eref{t:6::}
is always finite and includes the basic term (corresponding to the planet 
orbital period) and a few of its harmonics.

Let us note that we can examine characteristic features of the spectra of 
other processes (e.g. stellar pulsations) and compare them with the expansion 
of Keplerian induced radial-velocity variations. Such information might be 
crucial for the proper interpretation of observations.  

\section{Numerical Realization of Frequency Analysis}

Let ${\cal V}^0$ denote a set of radial velocities of a star obtained from
the Doppler-shift measurements. Using the least-squares method, we  fit the function 
$F_{1}(t)= A_1\sin(2\pi f_1 (t-t_0)) + B_1\sin(2\pi f_1 (t-t_0))$ to the data. 
As the first approximation of $f_1$, we take the frequency corresponding to the 
maximum of the Lomb-Scargle periodogram of ${\cal V}^0$ \cite[]{Lomb:76::,Scargle:82::,Press:92::}. 

At the $k$-th stage of this algorithm, we have a set of the residuals 
${\cal V}^k$ obtained by fitting function $F_{k}$ to the original observations, 
where
\begin{equation}
\label{m:1::}
F_{k}(t) = F_{k-1}(t) + A_k\sin(2\pi f_k (t-t_0)) + B_k\cos(2\pi f_k(t-t_0)). 
\end{equation}
This means that at the $k$-th stage of the algorithm we have the following
model function $F_k(t)$:
\begin{equation}
\label{m:2::}
F_k(t) = \sum_{j=1}^{k}A_j\sin(2\pi f_j (t-t_0)) + 
B_j\cos(2\pi f_j(t-t_0)).
\end{equation}
Then, using the periodogram of ${\cal V}^k$, we approximate $f_{k+1}$, and 
we fit function $F_{k+1}$ to the original set of radial-velocities. These steps 
are repeated  until the desired number of terms is obtained or until the final 
residuals are smaller than the assumed limit. We call the above algorithm  
{\em unconstrained 
frequency analysis}, as we assume that all frequencies we find are independent. 
However, we can modify our method to the constrained form. To this end, we assume that 
all terms have frequencies that are natural combinations of the chosen basic 
frequency $f_b$
\begin{equation}
\label{m:3::}
f_{j}= jf_{b}, \quad j=1,\ldots,k.
\end{equation}
Thus the model we fit to the observations has the form
\begin{equation}
\label{m:4::}
\displaystyle v_{r}(t) = \sum_{j=1}^{k}
A_j\sin(2\pi j f_b(t-t_0)) + 
B_j\cos(2\pi j f_b(t-t_0)).
\end{equation}
In this way parameters of the model are $f_{b},A_j,B_j$ and not $f_{j},A_j,B_j$;  
$j=1,\ldots,k$. We call this version of the algorithm the {\em constrained 
frequency analysis}. 

Numerical tests of this method can be found in \cite[]{Konacki:96::} and its 
successful applications to the PSR~B1257+12 pulsar timing observations 
in \cite[]{Maciejewski:97::,Konacki:97::}. In the next section we show how
this method works on a simulated set of radial velocity measurements.

\section{An Example - 70 Virginis}

For the purposes of this example, we have chosen the orbital parameters of 70 Vir  
\cite[]{Marcy:96b::}. They are presented in Table 1 where 
we also present several periods and corresponding amplitudes 
calculated from expansion \eref{t:6::}. For our tests, we 
computed  data that are very similar in  nature to the real observations. 
Thus we have 39 unevenly sampled  radial-velocities. We also added 
10 ms$^{-1}$ Gaussian noise resembling the real observational error.

In Figure \ref{fa70vir} we present results from the unconstrained frequency analysis for
the data described above. What can we learn about the planetary system
from these results? First, we notice that the eccentricity of the orbit
of 70 Vir is large since we are able to detect three periodic terms in the
data. In fact, from the ratio of the amplitudes of the basic term to its first 
harmonic, we can precisely calculate the eccentricity
\cite[see][]{Konacki:96::,Konacki:97::}. 

Let us assume now that we do not know that these radial-velocity variations are
of planetary origin. From the unconstrained frequency analysis for 70 Vir 
we find out that there are three periodic terms detectable, and, up to
the accuracy of the method, they are all natural combinations of a certain
basic frequency. This basic frequency corresponds to the term with the largest 
amplitude. It means that in fact, we see effects from the basic frequency and 
its two subsequent harmonics. Moreover, the ratio of their amplitudes resembles 
a process of planetary (Keplerian) origin. It is extremely unlikely that there 
are other reasonable processes that can produce such spectra and in this way 
mimic a planet. Thus, taking into account all the facts, the most natural 
answer would be that we observe radial-velocity variations induced by a 
planet in an eccentric orbit.

In summary,  frequency analysis can supply us with strong constraints
on the possible models of observed radial-velocity variations and verify
the orbital solution obtained from a least-squares fit. We should also 
notice that we can apply the constrained frequency analysis as well; it usually 
results in a better fit and can be used to obtain better approximations of 
values of the amplitudes and basic frequency. Finally, one can compare this
result with the behavior of $\chi^2$ in Figure \ref{70virplane}. The
comparison shows that in many cases the frequency analysis will not be superior 
to the least-squares fit but will give an idependent insight into the data. 
Therefore, it is appropriate to use both methods while analyzing radial velocity 
measuerements, as the frequency analysis is numerically not demanding
and has the capablility of confirming the least-squares' findings. 

We should also mention that it is possibible to interpret the observations in 
a different way which might be as well justified as the original assumption.
Namely, it is possible to show that in some cases a planetary system
consisting of two (or more) planets can be misinterpreted as one planet
in a highly eccentric orbit and vice versa. This is easily understandable 
since from the frequency analysis we know that an eccentric orbit just
produces a finite set of harmonics in the real data which subsequently 
might be interpreted in
different manners. As there are some difficulties with the explanation of
the existence of planets with highly eccentric orbits (see the paper of 
\cite{Marcy:98::a} and reference therein for a discussion of this topic)
and there is no obvious reason why the planets discovered around the Solar-type 
stars should be the only ones in their systems the above idea might be
worth checking.

\section{Frequency Analysis of 16 Cygni B observations}

We applied the frequency analysis to real observations 
of 16 Cygni B. According to \cite{Cochran:97::} the planet is believed to 
move in an highly eccentric orbit (e=0.634). We calculated theoretical 
amplitudes of the first harmonics of expansion \eref{t:6::}, presented in 
Table 2. For calculation we took the orbital parameters obtained by means of 
the least squares method by \cite{Cochran:97::}. Comparing the values of amplitudes 
from Table 2 and the mean error of observations we  concluded that we could detect 
at least the main frequency $n$ and its harmonic $2n$.  We performed the 
frequency analysis of the real observations and we were only able 
to detect the basic frequency corresponding to the orbital period 
(see Figure \ref{fa16cyg}). After the subtraction of this term, there were no
other dominant frequencies present in the data. 
Thus, our result obtained with the help of frequency analysis was inconsistent 
with  result  of \cite{Cochran:97::} who reported the eccentricity 0.634
with uncertainty 0.082. This is, however, consistent with the results from 
Section 2. Stricly speaking, the frequency analysis reveals
that the current quality of the data available for 16 Cygni B does not allow
us to determine the eccentricity of the orbit due to the absence of any
harmonics and confirms the findings based on the $\chi^2$ topology and
the bootstrap method for these observations.

The final conclusion is that the real error of the eccentricity
is much greater than that reported by the least-squares method, making
the determination of the  eccentricity of 16 Cygni B almost impossible. This
fact can be easily derived from the frequency analysis. The absence of
any harmonics just indicates that the eccentricity must remain 
undetermined. With the frequency analysis one gets this correct result
without the time consuming $\chi^2$ map or the bootstrap method.

\section{Conclusions}

In this paper we have shown that results obtained from least-squares
fits of Keplerian orbits to real Doppler-shift measurements may lead
to incorrect interpretations. Specifically, they may give unrealistic 
or even entirely false values of parameters and their uncertainities.   
In order to solve these problems we have proposed a new method, frequency
analysis, which efficiently provides an independent test of the reliability 
of determined orbital parameters. This method may deliver a substantial 
revision of the current values of planets' high eccentricities that are 
essential for our understanding of the formation and evolution of planetary 
systems. It might even lead to hints that some of the observed high eccentric 
planets are in fact planetary systems consisting of more than one planet
or at least provide an independent point of view on the same data.
These facts, together with the ease of applicability of frequency analysis, 
make our method worth trying on future observations if not for the data 
already gathered.    

\acknowledgments{This work was  supported by KBN grant 2P03D.023.10.}

\newpage

\newcommand{\noopsort}[1]{}

\newpage
%
%
%
%
\figcaption[f3.ps]
{\label{16cygbplan}$\chi^2$ map on the plane $(e,a)$ of the real data of 
16 Cygni B. The filled square indicates the global minimum that is 
different than the orbital solution found by Cochran {\it et al.} (1997) 
denoted by the filled star. The significance levels "$1\sigma$", "$2\sigma$" 
and "$3\sigma$" are presented.}

%
%
\figcaption[f5.ps]
{\label{bootstrap} Bootstrap estimation of the distribution of $(e,a)$
based on the sample consisting of $10^{4}$  synthetic sets of 
data ({\it left}). Distribution of $e$ ({\it right, top}) and $a$ 
({\it right, bottom}) calculated from the same sample. The distributions 
for both parameters are clearly not normal.}

%
%
%
%
\figcaption[f1.ps]
{\label{fa70vir} Unconstrained frequency analysis of the fake reflex
velocity of 70 Vir. The simulated data consist of 39 points  from 1988 
to 1996.25, 10 ms$^-1$ Gaussian noise was added. {\it a,b,c,d:}
Subsequent steps of the frequency analysis. Found periods and their amplitudes 
are presented. Numbers in parentheses are $3\sigma$ uncertainties of the 
parameters. There are three detectable periodicities in the data.
}

%
%
\figcaption[f4.ps]
{\label{70virplane}$\chi^2$ map on the plane $(e,a)$ of the fake data 
of 70 Virginis. The filled square indicates the global minimum found and 
the star indicates the parameters assumed while calculating the fake data. 
The significance level $3\sigma$ is presented.}

%
%
\figcaption[f2.ps]
{\label{fa16cyg}Unconstrained frequency analysis of the real reflex
velocity of 16 Cygni B. The data taken from the paper of 
Cochran {\it et al.} (1997) consist of 70 points from about 1988 to 1997. 
{\it a,b:} Two subsequent steps of the frequency analysis. As one can see, 
there is only one periodic term detectable ({\it a}) corresponding to the 
planetary period of about 800 days. After subtracting this periodicity 
there are no significant peaks in the data present. Solid and dash-dotted 
line correspond to 90 an 50 percent significance levels, respectively. }


\newpage
%
%
%
%
\begin{deluxetable}{ll|lll}
\tablewidth{0pt}
\tablecaption{Orbital and Theoretical Frequency Analysis Parameters of 70 Vir}
\startdata
\nl
\tableline
70 Vir \nl
\tableline
$P$ (days) & 116.67 & $k=1$& $P1=116^{\mbox{d}}.67$ & $A_{P1} = 261.36 \mbox{ ms}^{-1}$ \nl
$T_p$ (JD) &  2448990.403 & $k=2$ & $P2=58^{\mbox{d}}.335$ &  $A_{P2} = 101.08 \mbox{ ms}^{-1}$ \nl
$e$ & 0.40 & $k=3$ & $P3=38^{\mbox{d}}.89$ & $A_{P3} = 43.83 \mbox{ ms}^{-1}$\nl
$\omega$ (deg) & 2.1 & $k=4$  & $P4=29^{\mbox{d}}.1675$ & $A_{P4} = 19.99 \mbox{ ms}^{-1}$ \nl
$a_1\sin i$ (AU) & 0.00312 & $k=5$ & $P5=23^{\mbox{d}}.334$ & $A_{P5} = 9.39 \mbox{ ms}^{-1}$ \nl
\enddata
\end{deluxetable}
%
%
\begin{deluxetable}{ll|lll}
\tablewidth{0pt}
\tablecaption{Orbital and Theoretical Frequency Analysis Parameters of 16
Cyg B}
\startdata
\nl
\tableline
16 Cyg B \nl
\tableline
$P$ (days) & 800.8 & $k=1$& $P1=800^{\mbox{d}}.8$ & $A_{P1} = 28.89\mbox{ ms}^{-1}$ \nl
$T_p$ (JD) &  2448935.3 & $k=2$ & $P2=400^{\mbox{d}}.4$ &  $A_{P2} = 16.14\mbox{ ms}^{-1}$ \nl
$e$ & 0.634 & $k=3$ & $P3=200^{\mbox{d}}.2$ & $A_{P3} = 10.23\mbox{ ms}^{-1}$\nl
$\omega$ (deg) & 83.2 &   &  \nl
$K$ ($\mbox{ms}^{-1}$) & 43.9 &  \nl
\enddata
\end{deluxetable}
%
%
%
\newpage
\begin{figure}
\figurenum{1}
\epsscale{0.8}
\plotone{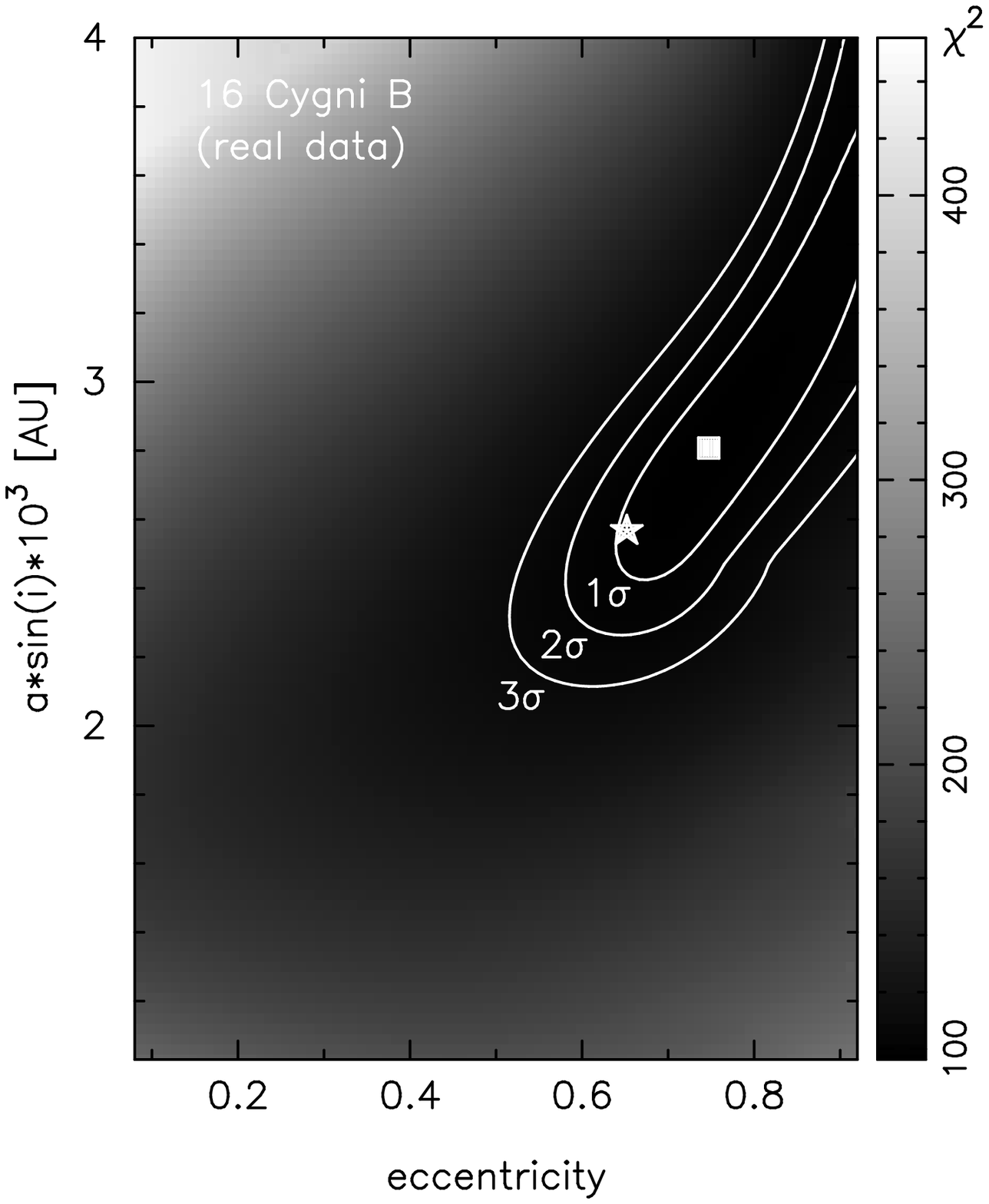}
\caption{}
\end{figure}
\begin{figure}
\figurenum{2}
\epsscale{0.8}
\plotone{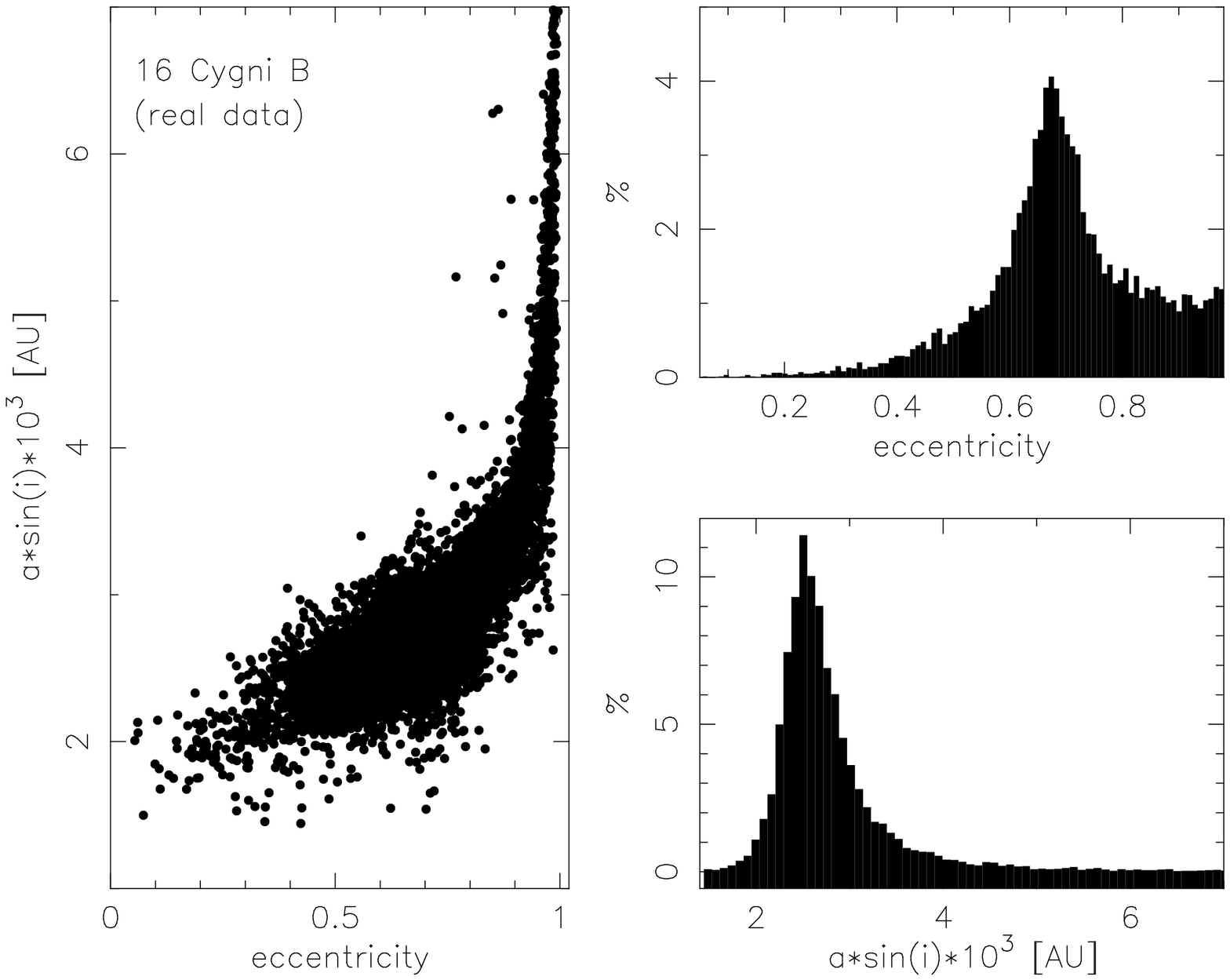}
\caption{}
\end{figure}
%
%
\begin{figure}
\figurenum{3}
\epsscale{0.7}
\plotone{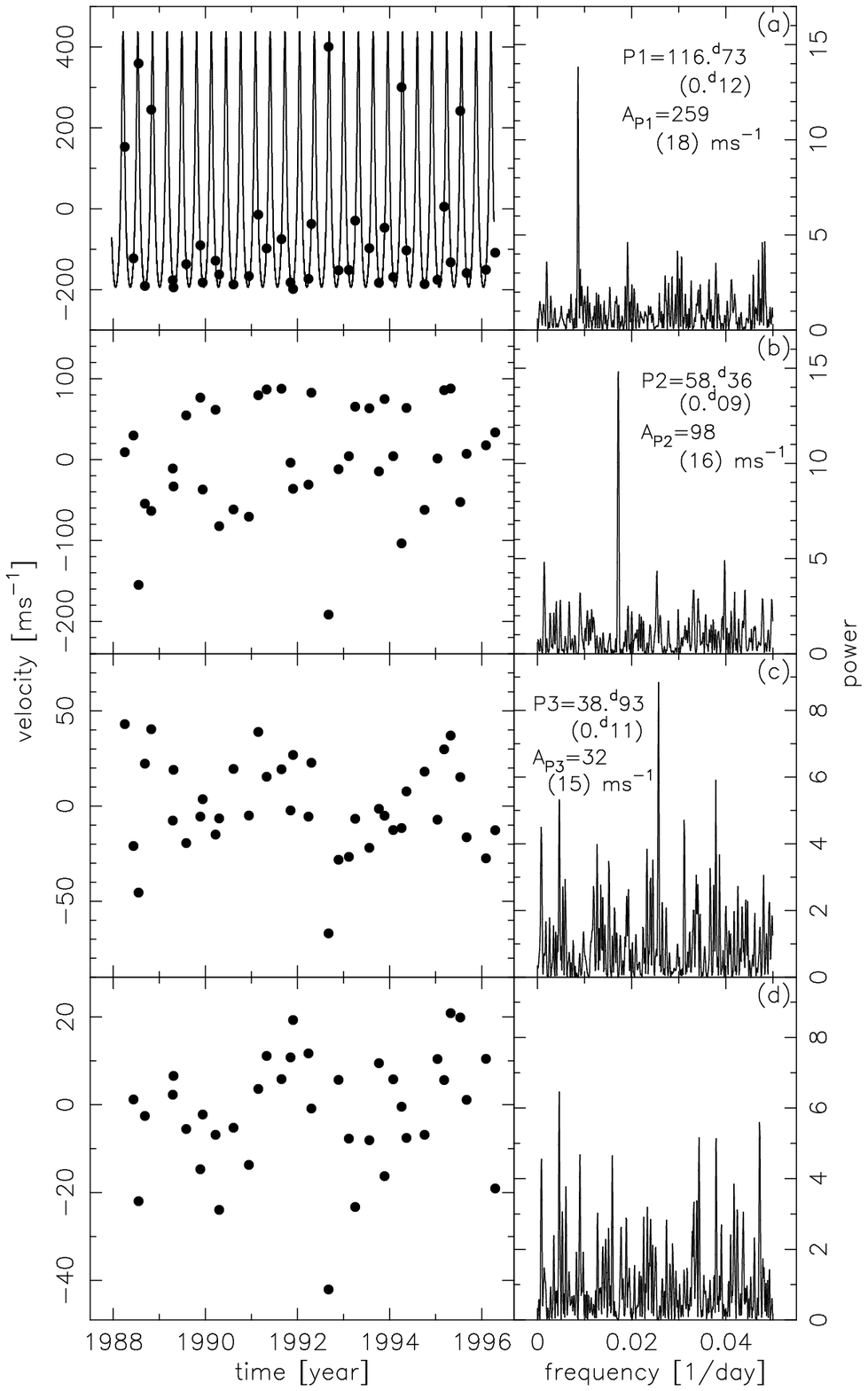}
\caption{}
\end{figure}
%
\begin{figure}
\figurenum{4}
\epsscale{0.8}
\plotone{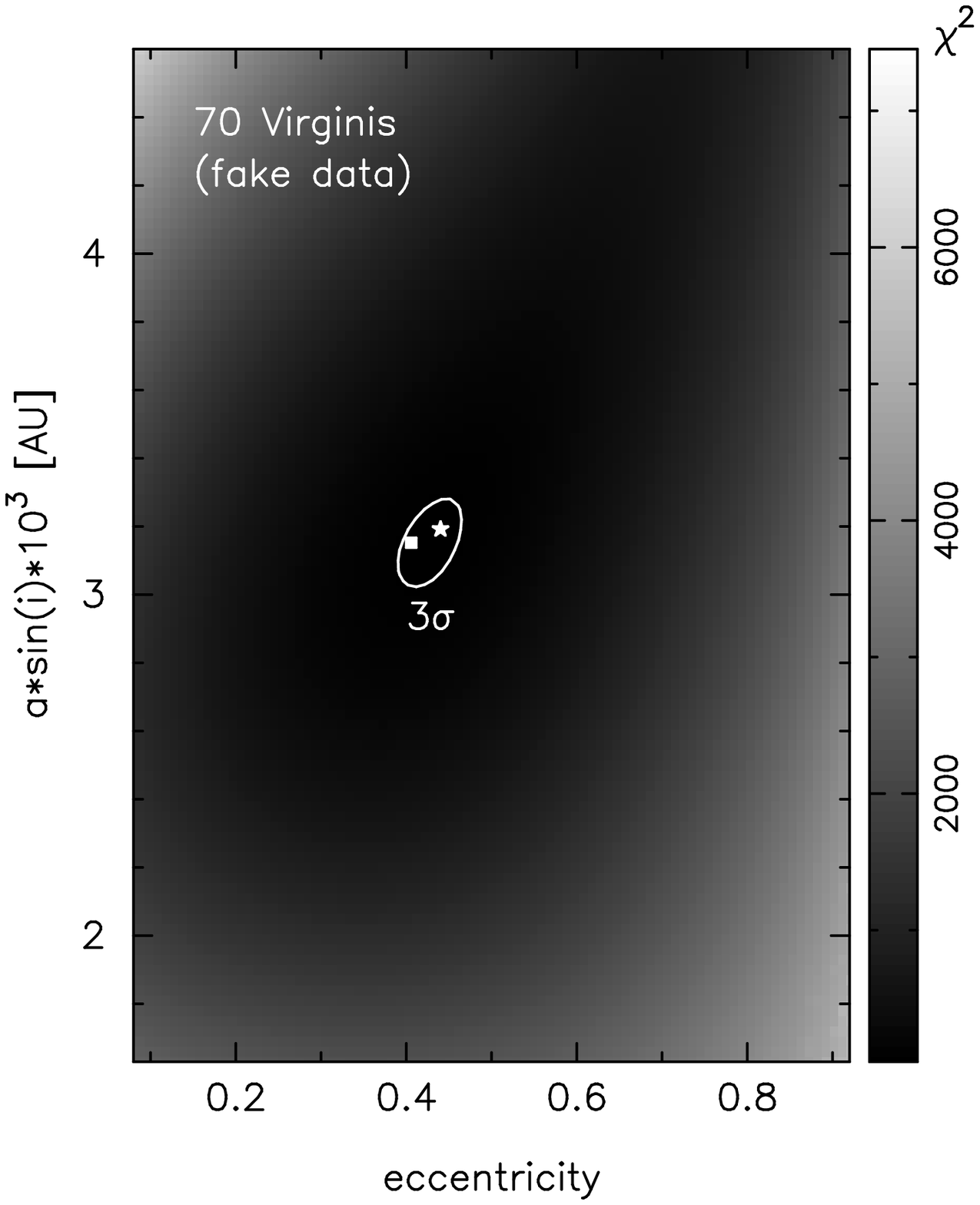}
\caption{}
\end{figure}
%
%
\begin{figure}
\figurenum{5}
\epsscale{0.8}
\plotone{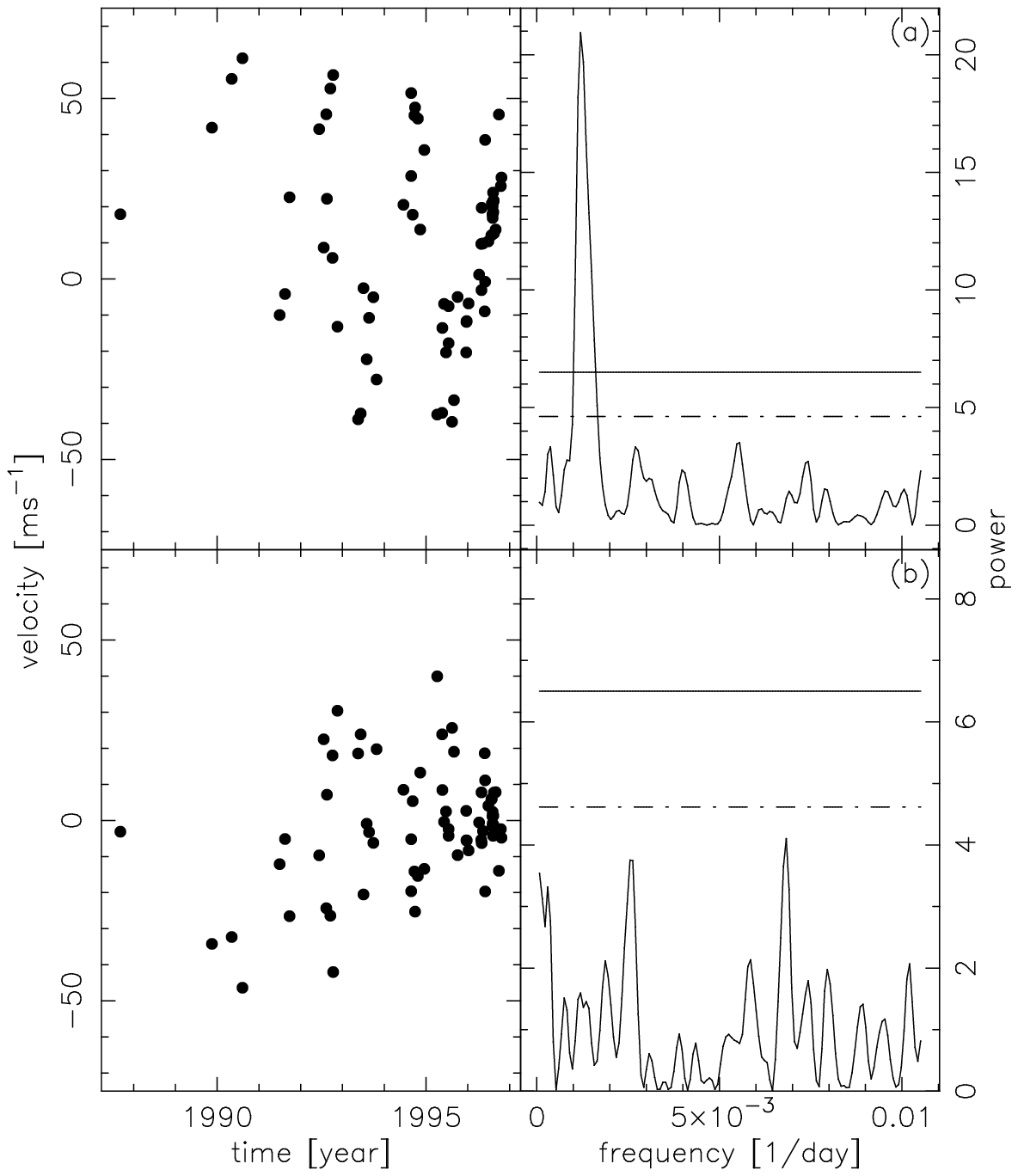}
\caption{}
\end{figure}

\end{document}